\documentclass [%
superscriptaddress,
preprint,
 amsmath,amssymb,
 showkeys
]{revtex4-1}

\usepackage{graphicx}
\usepackage{dcolumn}
\usepackage{bm}
\usepackage{color}
\usepackage{float}
\usepackage[utf8]{inputenc}

\usepackage{multirow}
\usepackage{hyperref}
\usepackage{amsmath}
\usepackage{xr}

\usepackage{soul,xcolor}
\soulregister\cite7
\soulregister\ref7
\setstcolor{blue}
\begin{document}

\title{Upper limit on nonlinear optical processes: shift current and second harmonic generation in extended systems}

\author{Liang Z. Tan}
\affiliation{Department of Chemistry, University of Pennsylvania, Philadelphia, Pennsylvania 19104, USA}
\author{Andrew M. Rappe}
\affiliation{Department of Chemistry, University of Pennsylvania, Philadelphia, Pennsylvania 19104, USA}

\date{\today}



\maketitle

\textbf{The response functions of a material characterize its behavior under external stimuli, such as electromagnetic radiation. Such responses may grow linearly with the amplitude of the incident radiation, as is the case of absorption, or may be nonlinear. The latter category includes a diverse set of phenomena such as second harmonic generation (SHG)~\cite{Prasad91}, shift current~\cite{Fridkin73p71,vonBaltz81p5590,Belinicher80p199,Tan16p16026}, sum frequency generation~\cite{Shen89p519}, and excited state absorption~\cite{Bellier12p15299}, among others. Despite decades of research into nonlinear response theory, and the occasional discovery of materials with large nonlinear responses, there has been no systematic investigation into the maximum amount of nonlinear optical response attainable in solid-state materials. In this work, we present an upper bound on the second-order response functions of materials, which controls the SHG and shift current responses. We show that this bound depends on the band gap, band width, and geometrical properties of the material in question. We find that Kuzyk's bound~\cite{Kuzyk00p1218} for the maximum SHG of isolated molecules can be exceeded by conjugation or condensation of molecules to form molecular solids, and that strongly coupled systems generally have larger responses than weakly coupled or isolated ones~\cite{Zhou06p2891,Szafruga10p379,Perez-moreno09p5084,Coe10p3496}. As a proof of principle, we perform first-principles calculations of the response tensors of a wide variety of materials, finding that the materials in our database do not yet saturate the upper bound. This suggests that new large SHG and shift current materials will likely be discovered by future materials research guided by the factors mentioned in this work. }

The shift current bulk photovoltaic effect, which is the generation of current in a bulk single-phase material under illumination, and second harmonic generation, which is the doubling of the frequency of incident light, are both second order nonlinear optical effects. The induced current ($J$) is proportional to the second power of the electric field ($E$) of the incident light,

\begin{equation}
J_r(\omega_{\text{out}}=0)=\sigma^{\text{SC}}_{rst}
(\omega_{\text{in}})
E_s(\omega_{\text{in}}) E_t(-\omega_{\text{in}})   
\end{equation}

\noindent
for the shift current, and

\begin{equation}
J_r(\omega_{\text{out}}=2\omega_{\text{in}})=\sigma^{\text{SHG}}_{rst}
(\omega_{\text{in}})
E_s(\omega_{\text{in}}) E_t(\omega_{\text{in}}) 
\end{equation}

\noindent
for the SHG, where $\omega_{\text{in}}$ is the frequency of the incident light, and $\omega_{\text{out}}$ is the frequency of the response. As a result~\cite{vonBaltz81p5590,Belinicher80p199}, both the shift current and the SHG are present only in materials lacking inversion symmetry. The shift photocurrent can therefore be generated without the need for a traditional $p$-$n$ junction, which has motivated the field of ferroelectric photovoltaics~\cite{Fridkin73p71,Huang10p134,Yuan14p6027,Tan16p16026,Paillard16p5153}.

In extended systems, the second order perturbation theory expressions for the shift current and SHG second order conductivities are~\cite{vonBaltz81p5590, Sipe00p5337} 

\begin{equation}
\label{eq:sc}
\begin{aligned}
\sigma^{\text{SC}}_{rst}(\omega_{\text{in}}) = \pi e  \left (\frac{e}{m\hbar\omega_{\text{in}}} \right )^2
\sum_{cvk} &
\langle c \lvert p_r \rvert v \rangle 
\langle v \lvert p_s \rvert c \rangle \\
&\quad\delta(\omega_c - \omega_v - \omega_{\text{in}}) \mathcal{R}_{rt}(c,v,k)
\end{aligned}
\end{equation}

\begin{equation}
\label{eq:shg}
\begin{aligned}
\sigma^{\text{SHG}}_{rst}(\omega_{\text{in}}) = \pi e  \left (\frac{e}{m\hbar\omega_{\text{in}}} \right )^2
\sum_{cvk} &
\langle c \lvert p_r \rvert v \rangle 
\langle v \lvert p_s \rvert c \rangle \\
&\quad(-\delta(\omega_c - \omega_v - \omega_{\text{in}}) +\frac{1}{2} \delta(\omega_c - \omega_v - 2\omega_{\text{in}})) \mathcal{R}_{rt}(c,v,k)
\end{aligned}
\end{equation}

\noindent
Here, the sum over states includes all conduction ($c$) and valence ($v$) bands and corresponding integrals over the Brillouin zone. The components of the momentum operator are denoted by $p_r$. The shift vector

\begin{equation}
\label{sv}
\mathcal{R}_{rt}(c,v,k) = -\frac{\partial}{\partial k_t} \arg \langle c \lvert p_r \rvert v \rangle - [\chi_{vt}(k) - \chi_{ct}(k)]
\end{equation}

\noindent contains the Berry connections ($\chi$), and has been linked to topological ideas in nonlinear optics~\cite{Morimoto16pe1501524,Tan16p237402,Nagaosa17p1603345}. The shift vector $\mathcal{R}$ can be understood as a generalized gauge invariant $k$-space derivative of the $p$ operator \cite{Aversa95p14636,Sipe00p5337}, and it is odd under the interchange of $c$ and $v$ bands. This formalism has been succussfully used in first-principles calculations of shift current and SHG~\cite{Hughes96p10751,Young12p116601,Tan16p16026}. 

In applications of SC and SHG, the quantity of interest is often not the the value of the response function at a fixed frequency, but rather the values it takes across a range of frequencies. For instance, the total current produced by a photovoltaic device is given by the integral of $\sigma(\omega_{\text{in}})$ weighted by the radiation intensities at all incident frequencies. Alternatively, one may be interested in the average SHG response of a material across a frequency range instead of some predetermined frequency. We therefore propose the integral $M = \lvert \int \, \sigma dE \rvert$ as a metric for evaluating the overall magnitude of the nonlinear response of a material, where $E=\hbar \omega$. For the frequency range of this integral, we consider contributions from the lowest conduction and highest valence bands of the material. This is therefore a metric for the lower frequency range of the nonlinear optical spectrum of a material. 
Despite the truncation of Eqs.~\ref{eq:sc}, \ref{eq:shg} to two bands, it should be stressed that the bounds derived below are not the bounds of a purely two-level model system, but are bounds for the lowest two levels of a multi-level system. The difference is that the second-order susceptibility for a pure two-level system vanishes~\cite{Kuzyk13p297}, whereas higher energy bands are taken into account even in the two lowest levels of Eqs.~\ref{eq:sc}, \ref{eq:shg} via the application of a sum rule~\cite{Sipe00p5337}.   

We therefore consider the quantity (including a factor of 2 for spin degeneracy of bands)

\begin{equation}\label{eq:metric}
M =  \frac{2\pi e^3}{m^2\hbar\omega_{\text{in}}^2} 
\left \lvert
\sum_{k} 
\langle c \lvert p_r \rvert v \rangle 
\langle v \lvert p_s \rvert c \rangle 
\mathcal{R}_{rt}(c,v,k) 
\right \rvert
\end{equation}

\noindent as a measure of the overall magnitude of shift current or SHG responses, keeping in mind that $M^{\text{SC}} = M$, and $M^{\text{SHG}} = M/2$, with the additional factor of $1/2$ arising from Eq.~\ref{eq:shg}. SHG is often measured in terms of the nonlinear susceptibility, which is related to the nonlinear conductivity by $\chi^{(2)}=\sigma^{\text{SHG}}/(2i\omega\epsilon_0)$.

We begin our derivation of an upper bound on $M$ by considering the Hamiltonian of the $c$ and $v$ bands, which determines the quantities appearing in Eq.~\ref{eq:metric}. A generic Hamiltonian for this two band system (which may be obtained, for instance, through the use of maximally localized Wannier functions~\cite{Marzari97p12847,Marzari12p1419}) can be written as

\begin{equation}\label{eq:ham}
H(k) = \vec{h}(\vec{k}) \cdot \vec{\tau} = \sum_{i=1}^3 h_i(k) \tau_i
\end{equation}

\noindent
where the $\tau_i$ are Pauli matrices representing the band degree of freedom. The shift current of such a  Hamiltonian was derived in~\cite{Fregoso16preprint}. For simplicity, we focus here on the longitudinal tensor components of the nonlinear response functions, $\sigma_{iii}$ along some direction $\vec{v}$. With the above assumptions, our metric for the overall shift current magnitude becomes

\begin{equation}\label{eq:hhh}
M = \frac{\pi e^3}{2\hbar} \left \lvert
\int \frac{d^3k}{(2\pi)^3} 
\frac{\vec{h}(\vec{k}) \cdot \vec{h}'(\vec{k}) \times \vec{h}''(\vec{k})}{E(\vec{k})^3} 
\right \rvert
\end{equation}

\noindent
where the derivatives $\vec{h}'=\frac{d}{dk}\vec{h}$, $\vec{h}''=\frac{d^2}{dk^2}\vec{h}$ are taken along the direction of light polarization and current $\vec{v}$, and $E(\vec{k})=\lvert \vec{h}(\vec{k}) \rvert$ is the band transition energy at $\vec{k}$. From the appearance of $E(\vec{k})$ in the denominator of this expression, it can already be seen that small band gaps tend to favor large nonlinear responses, as has been noted in ~\cite{Cook17p14176}. This, however, does not mean that minimizing the band energy throughout the entire Brillouin zone would yield the greatest possible response, because of the competing factors of $\vec{h}'$ and $\vec{h}''$ in the numerator, which favor variation in the Hamiltonian. In other words, dispersive bands would also tend to increase the amount of response. We therefore expect that a balance of these two factors determines the amount of response. 

In a system with a fixed band gap, a rescaling of the band width will increase the value of $\vec{h}''$, and hence of $M$, without bound. In real materials, the Hamiltonian is restricted to physically attainable values. In the tight-binding picture, the band width grows with the strength of the hopping between atomic sites. We therefore impose the restriction that the Fourier components of $\vec{h}$, which are the hopping amplitudes between Wannier functions~\cite{Marzari97p12847}, are bounded in magnitude and decay exponentially with distance 

\begin{equation}\label{eq:expdecay}
\begin{aligned}
\vec{h}(\vec{k}) =& \sum_{n_1n_2n_3} \vec{h}_{n_1n_2n_3} \exp\left(i\vec{k}\cdot(n_1\vec{R}_1+n_2\vec{R}_2+n_3\vec{R}_3)\right)\\
\lvert \vec{h}_{n_1n_2n_3} \rvert  <& \,A \, \exp \left (-\frac{n_1}{\xi_1}-\frac{n_2}{\xi_2}-\frac{n_3}{\xi_3} \right)
\end{aligned}
\end{equation}

\noindent Here, $A$ is the overall scale for the magnitude of the Hamiltonian and $\xi_i$ are the hopping ranges which can be different along different lattice directions $\vec{R}_i$. Since $\vec{h}'$ and $\vec{h}''$ scale with $A$, and transitions $E(\vec{k})$ are no less than the band gap $E_g$, the form of Eq.~\ref{eq:hhh} suggests that an upper bound for $M$ is proportional to $(A/E_g)^2$. We show (methods section~\ref{sec:genbound}) that this is indeed the case, with $M$ obeying the general bound

\begin{equation}\label{eq:genbound}
M <  \frac{\pi e^3}{2\hbar}  \left ( \frac{A}{E_g} \right)^2 \Xi\left(\overset{{\tiny\leftrightarrow}}{R},\vec{\xi},\vec{v} \right)
\end{equation}

\noindent
where $\Xi\left(\overset{{\tiny\leftrightarrow}}{R},\vec{\xi},\vec{v} \right)$ is a dimensionless geometrical factor depending on the crystal lattice ($\overset{{\tiny\leftrightarrow}}{R}=(\vec{R}_1,\vec{R}_3,\vec{R}_3)$), hopping ranges ($\vec{\xi}$), and the measurement direction $\vec{v}$. This bound holds for all non-zero values of $A$
, $E_g$, and $\xi$. The detailed form of the geometrical factor is
\begin{equation}\label{eq:geomfac}
\begin{aligned}
\Xi\left(\overset{{\tiny\leftrightarrow}}{R},\vec{\xi},\vec{v} \right) = \frac{1}{V}
\biggl[
\sum_{i,j,l=(1\;2\;3)} & \tilde{v}_i 
\frac{2e^{-1/\xi_i}}{(1-e^{-1/\xi_i})^2}  
\frac{1+e^{-1/\xi_j}}{1-e^{-1/\xi_j}}
\frac{1+e^{-1/\xi_l}}{1-e^{-1/\xi_l}}
\biggr] \cdot\\
\biggl[
\sum_{i,j,l=(1\;2\;3)} & \tilde{v}_i^2 
\frac{2(e^{-1/\xi_i}+e^{-2/\xi_i})}{(1-e^{-1/\xi_i})^3}
\frac{1+e^{-1/\xi_j}}{1-e^{-1/\xi_j}}
\frac{1+e^{-1/\xi_l}}{1-e^{-1/\xi_l}} \\
+ & 2 \tilde{v}_i\tilde{v}_j
\frac{2e^{-1/\xi_i}}{(1-e^{-1/\xi_i})^2}  
\frac{2e^{-1/\xi_j}}{(1-e^{-1/\xi_j})^2} 
\frac{1+e^{-1/\xi_l}}{1-e^{-1/\xi_l}}
\biggr] 
\end{aligned}
\end{equation}

\begin{figure}
\centering
\includegraphics[width=\textwidth]{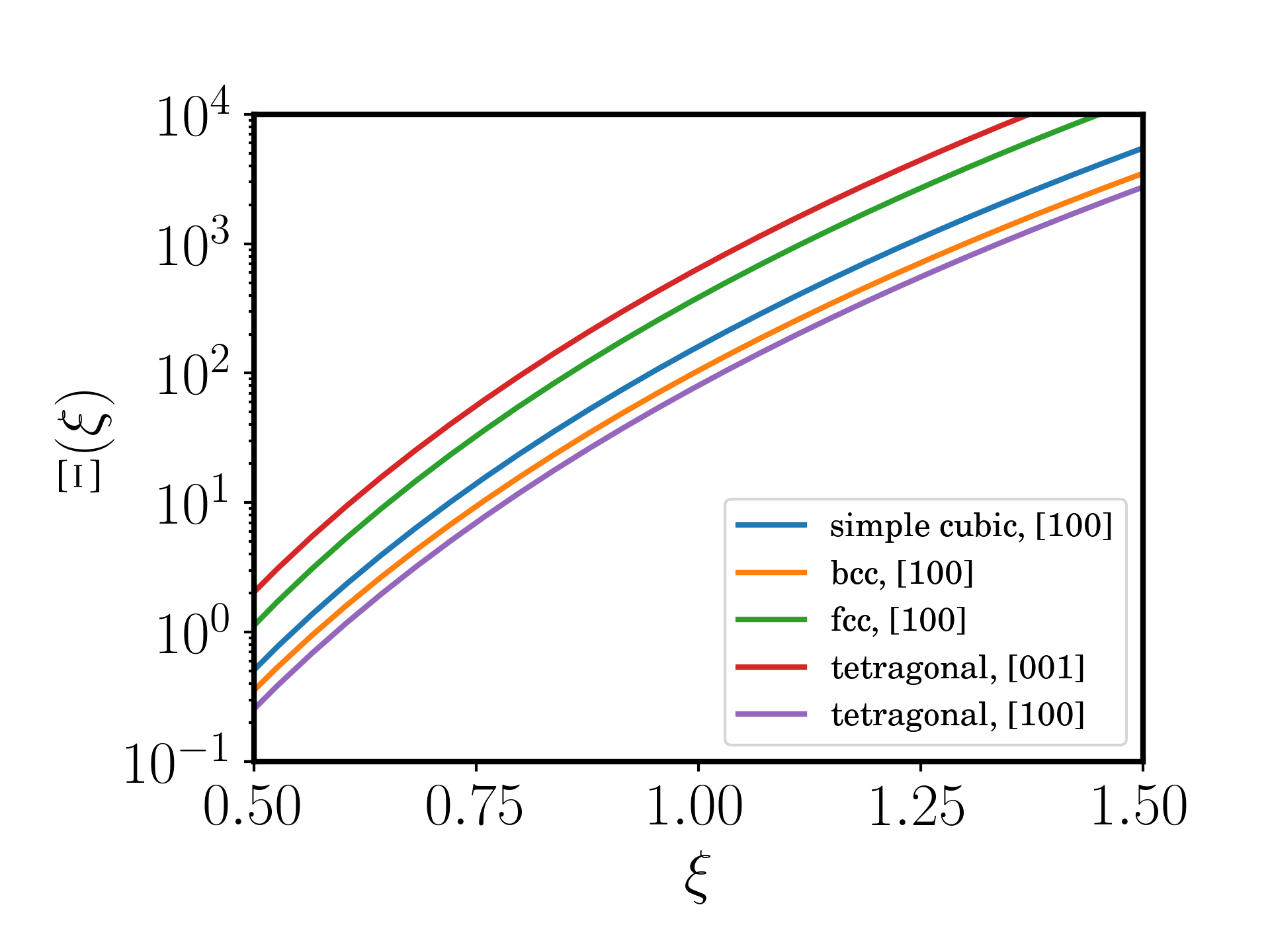}
\caption{
Geometrical factor $\Xi(\xi)$ for the upper bound on nonlinear optical response, as a function of the hopping range $\xi$, defined in Eq.~\ref{eq:geomfac} of the text. The geometrical factor is shown for different lattices, and for different measurement directions. Anisotropic lattices show the highest potential for large nonlinear responses. Here, the tetragonal lattice has $c/a=2.0$ ratio, and has largest nonlinear response upper limit for light polarization and current measurement directions along the $c$-axis. 
}
\label{fig:geomfac}
\end{figure}

\noindent Here, the volume of the unit cell is $V=\lvert\det R\rvert$, the sums run over all cyclic permutations of $(1\;2\;3)$, and $\tilde{v}_i = \sum_j R_{ji} v_j$. This geometrical factor is an increasing function of the hopping ranges $\vec{\xi}$. The geometrical factor can vary greatly in magnitude depending on the range of hopping (Fig.~\ref{fig:geomfac}), with the exact value depending on the shape of the unit cell and direction of measurement. $\Xi$ is a rapidly growing function of the hopping ranges, asymptoting to $\xi^9$ for large values of $\xi$. We therefore expect this factor to be large in materials where second- or higher-neighbor hopping is comparable to nearest neighbor hopping. Previous studies~\cite{Tan16p16026} have noted, based on observing trends in the calculated shift current across materials classes, that highly covalent materials with delocalized wavefunctions tend to have large shift currents. In the context of Eq.~\ref{eq:genbound}, we recognize two distinct, but related reasons for this trend. Firstly, materials with strong covalent bonds would have large $A$ values. Secondly, even if a material does not have particularly strong first-neighbor hopping, the presence of further-neighbor hopping comparable in magnitude would tend to delocalize wavefunctions and increase the geometrical factor $\Xi$. A further examination of Fig.~\ref{fig:geomfac} shows that highly asymmetric unit cells tend to increase $\Xi$ as well, which is supported by the observation that some of the materials with highest predicted shift current contain 1-dimensional chains or motifs~\cite{Brehm14p204704,Liu17p6500}.

\begin{figure}
\centering
\includegraphics[width=\textwidth]{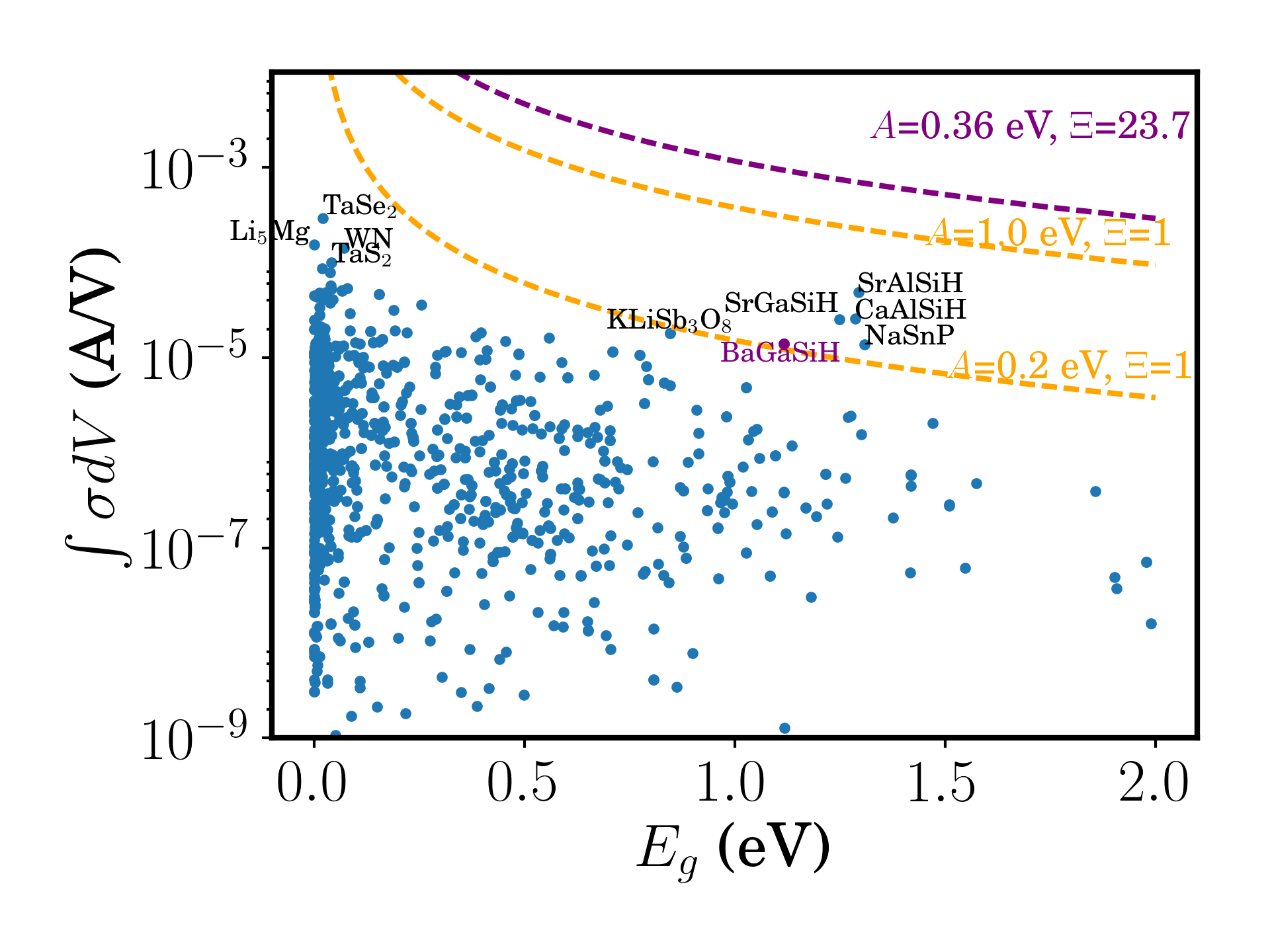}
\caption{
Integrated nonlinear response for a test set of semiconductors and semimetals. The largest tensor component of the integrated nonlinear response for each material is plotted against the band gap. Dashed lines indicate the value of the nonlinear response upper bound (Eq.~\ref{eq:genbound}) as a function of the band gap, for different values of hopping strength ($A$) and geometrical factor ($\Xi$). Select materials with large responses (Li$_5$Mg, TaSe$_2$, TaS$_2$, WN), or which deviate from the overall trend (KLiSb$_3$O$_8$, SrGaSiH, BaGaSiH, SrAlSiH, CaAlSiH, NaSnP) are indicated on the plot. Shown in purple are the integrated nonlinear response of BaGaSiH and the theoretical bound constructed using the $A$ and $\Xi$ values of BaGaSiH. 
}
\label{fig:abinit}
\end{figure}

We compare our derived bound Eq.~\ref{eq:genbound} with ab-initio calculations in Fig.~\ref{fig:abinit}. We have selected a test set of 1246 non-centrosymmetric materials from the \textsc{Materials Project} database~\cite{Jain13p011002}, choosing those with less than 30 atoms per unit cell, for computational efficiency. We have restricted our calculations to non-magnetic materials, and to thermodynamically stable or metastable materials with decomposition energy of less than 0.1 eV/atom. These calculations were done with the PBE density functional~\cite{Perdew96p3865}, using norm-conserving RRKJ pseudopotentials~\cite{Rappe90p1227}, and using a planewave basis set with kinetic energy cutoff of 60 Ry. Spin-orbit coupling was included at the fully-relativistic level for all calculations. A Monkhorst-Pack
8$\times$8$\times$8 $k$-point mesh was used for the self-consistent evaluation of the charge densities and calculation of the nonlinear response tensors.

To make a direct comparison of ab-initio calculated response tensors to our analytical bound, we integrate the response tensors over an energy range corresponding to transitions between the lowest conduction band and highest valence band only. In  Fig.~\ref{fig:abinit}, we plot, for each material, the largest tensor component of $\lvert \int \, \sigma dE \rvert$. Superimposed on the figure are contours corresponding to values of the bound (Eq.~\ref{eq:genbound}) at particular values of $A$ and $\Xi$, with $E_g$ allowed to vary. Most of the materials in the database fall below the contour with $A$=0.2 eV and $\Xi=1$. We note that the trend of the ab-initio data-points closely tracks the shape of the contours, with the materials with the largest responses having the smallest band gaps. Among these materials are the semimetals TaSe$_2$, TaS$_2$, WN, and Li$_5$Mg. The experimentally measured SHG response of the Weyl semimetal TaAs was shown to be an order of magnitude larger than most other SHG materials~\cite{Wu16p350}. The converse, however, is not true: having a small band gap does not necessarily mean that a material has large nonlinear response, as can be seen from Fig.~\ref{fig:abinit}. More generally, Eq.~\ref{eq:genbound} is an upper bound rather than a correlation across the space of all materials.  

\begin{figure}
\centering
\includegraphics[width=\textwidth]{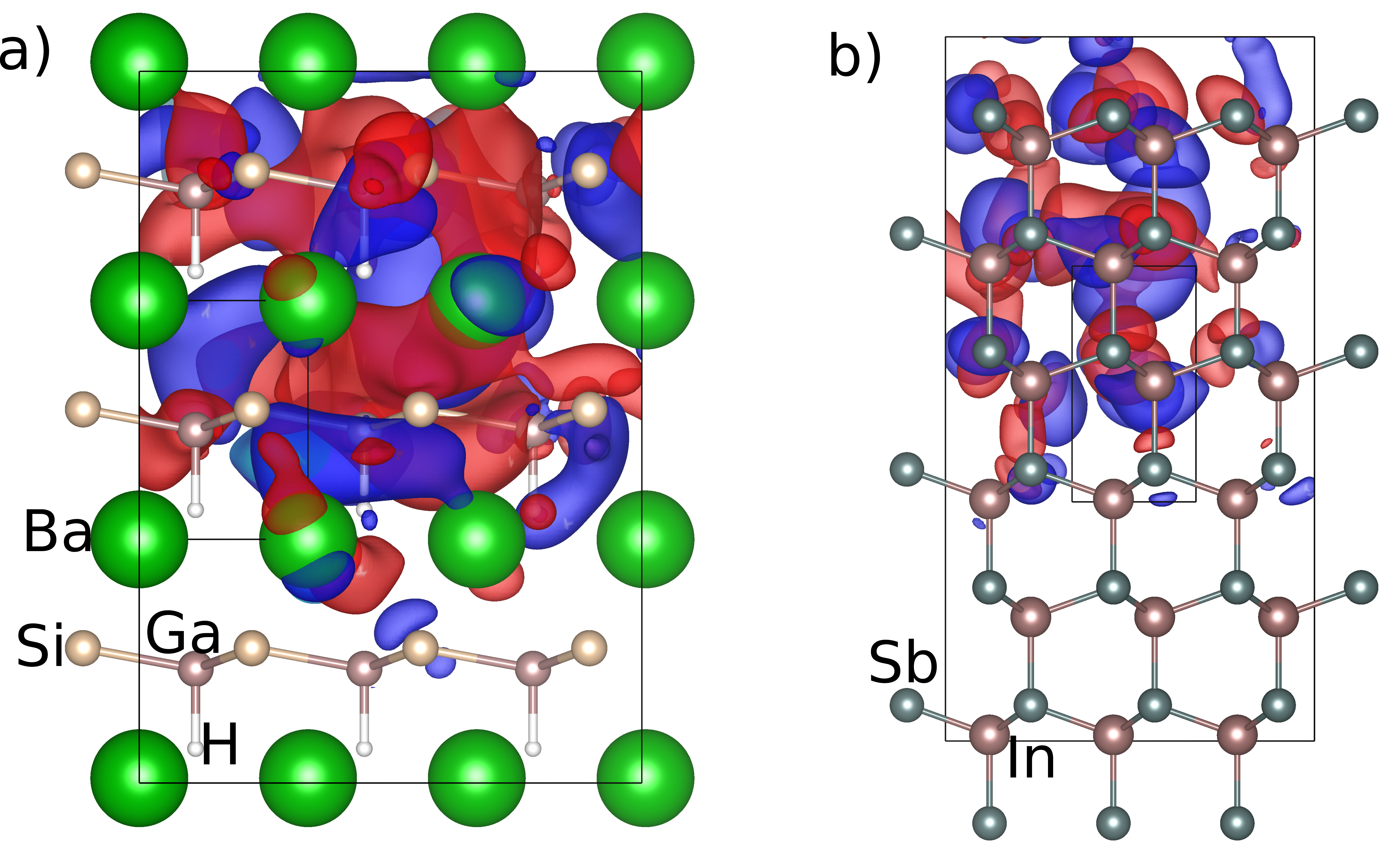}
\caption{
Wannier functions constructed from frontier (conduction and valence) orbitals of (a) BaGaSiH and (b) InSb. BaGaSiH is a large second harmonic generation and shift current material, with integrated response tensor (see text) $\int \sigma d\omega = 1.4 \times 10^{-5} A/V$. In contrast, InSb has low nonlinear response, with $\int \sigma d\omega = 3.1 \times 10^{-7} A/V$. These differences are explained in terms of the bonding character between the two materials. Isosurfaces of the Wannier functions of the two materials are plotted, with the isolevel chosen at 14\% of the maximum value of the Wannier function. The Wannier orbitals of BaGaSiH ($\xi=0.61$, $\Xi=23.7$) are more diffuse than those of InSb ($\xi=0.13$, $\Xi=0.38$), giving rise to longer range hopping in BaGaSiH. 
}
\label{fig:wan}
\end{figure}

In Fig.~\ref{fig:abinit}, there is a group of outliers which lie above the $A$=0.2 eV,$\Xi=1$ contour. This group contains several Zintl-type materials (AXYH, with A= group 1 or 2; X,Y=group 13--16). These materials are likely to have stronger or longer range bonding than other materials with the same band gap, and warrant further study into their photophysical properties. Among this group is BaGaSiH, with an integrated conductivity of $M=1.4 \times 10^{-5}$ A/V.  We have constructed maximally localized Wannier orbitals from its frontier conduction and valence bands, and fitted the hopping parameters of the resulting Hamiltonian (Eq.~\ref{eq:ham}) to an exponential dependence (Eq.~\ref{eq:expdecay}), obtaining values of $A=0.36$ eV, $\xi=0.61$, $\Xi=23.7$. The bound curve corresponding to the values of $A$ and $\Xi$ of BaGaSiH is shown in Fig.~\ref{fig:abinit}, indicating that the actual nonlinear response of BaGaSiH lies about two orders of magnitude below its theoretical bound. We compare this with a material with a relatively low amount of nonlinear response, InSb, with $M=3.1 \times 10^{-7}$ A/V. InSb has a comparatively less delocalized bonding, with $A=0.30$ eV, $\xi=0.13$, $\Xi=0.38$, and also seen in the smaller spatial extent of its Wannier orbitals (Fig.~\ref{fig:wan}).

We now consider the behavior of the bound in some limiting cases, to better understand the effects of localization and hopping. Eq.~\ref{eq:genbound} shows that wide band systems, which must necessarily have large $A$ values, have the potential to have large responses. The opposite limit of isolated systems (clusters or molecules), however, is not directly addressed by Eq.~\ref{eq:genbound} because such systems can have large or small $A$ values depending on the strength of the hopping within the isolated system. Nevertheless, a different upper bound can be derived for the nonlinear response of a system in the isolated limit. 
At this point, we distinguish between the phenomena of shift current and SHG. While SHG is regularly observed in molecules, the total steady state photocurrent in a system completely isolated from its surroundings must be zero. The total photocurrent is in fact a sum of different components, including the shift current and the recombination current. While the shift current is argued~\cite{Kral00p4851,Young12p116601,Tan16p16026} to be the dominant component in extended systems, it is cancelled by the recombination current in isolated systems. The following bound therefore applies to the SHG, or the shift current component of the total photocurrent in isolated systems.

\begin{figure}
\centering
\includegraphics[width=0.8\textwidth]{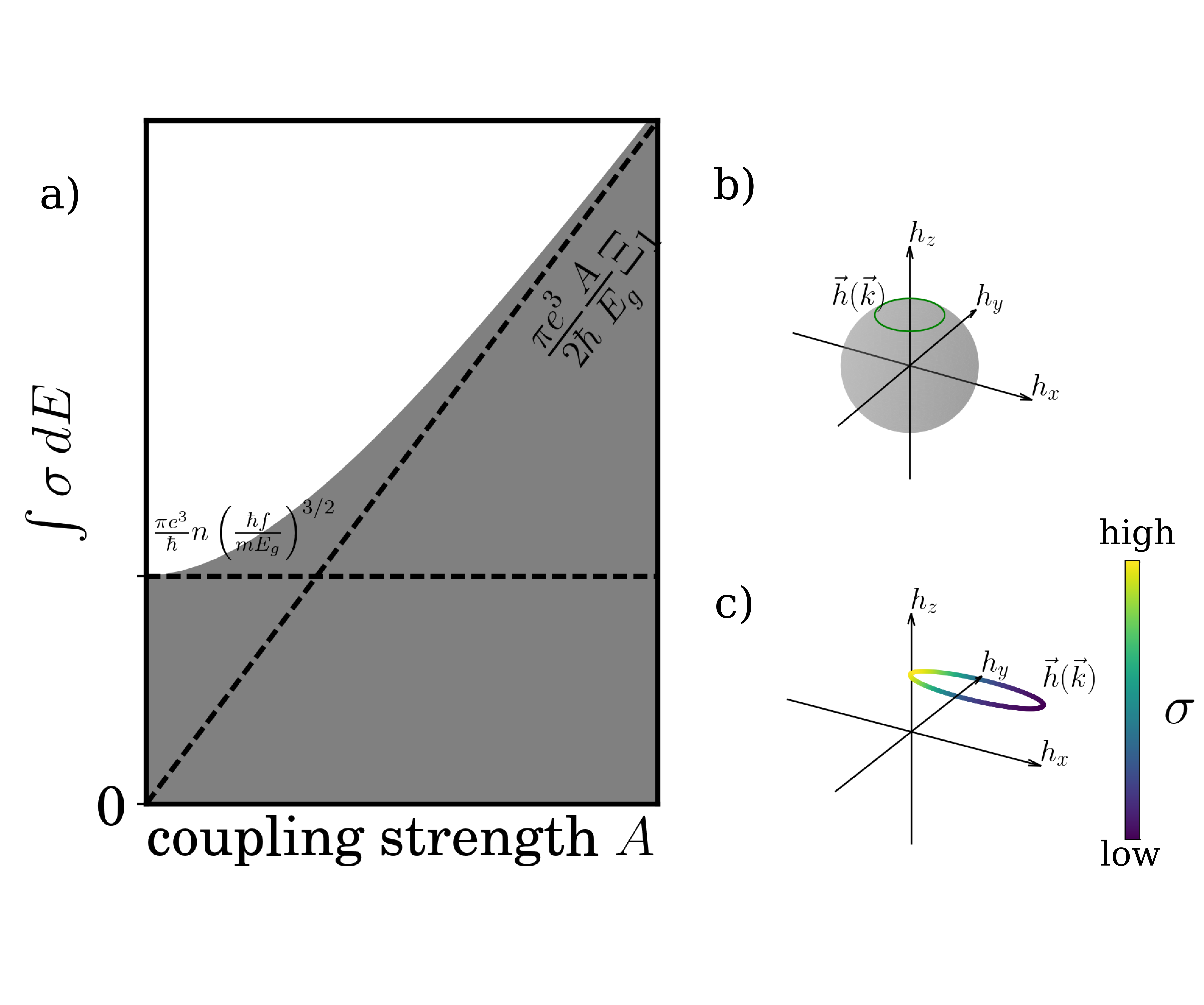}
\caption{
a) Schematic of nonlinear response bound as a function of hopping strength, for one-dimensional systems, as described by Eqs.~\ref{eq:molbound} and \ref{eq:1dbound}. The low hopping strength limit corresponds to the case of molecular, or isolated systems. In this limit, the maximum nonlinear response does not depend on hopping strength, at fixed oscillator strength and band gap. In comparison, the maximum nonlinear response for the high hopping strength limit grows linearly with hopping. The asymptotic bounds for these two limiting cases are plotted as dashed lines, while the shaded region denotes the permissible values of nonlinear response obtained by interpolating between these two limits.
b) Graph of the Hamiltonian components $\vec{h}(k)$, for the HOMO and LUMO of an isolated system. The constant energy splitting between HOMO and LUMO forces $\vec{h}(k)$ to lie on a sphere (shaded gray). The contribution to the nonlinear response (in color, green) is constant for all $k$.
c) Graph of the Hamiltonian components $\vec{h}(k)$, for the conduction and valence bands of an extended system. For this system, the band gap is located at $h_x=0,h_y=0$. The contribution to the nonlinear response is depicted in color, with the regions near the band gap having large (light color) contributions, and the regions away from the band gap having low (dark color) contributions.
}
\label{fig:bound1d}
\end{figure}

We use periodic boundary conditions, with multiple images of an isolated system arranged along the measurement direction, $\vec{v}$, and pass to the isolated limit by letting the hopping between different images go to zero. The effective Hamiltonian (Eq.~\ref{eq:ham}) of this system takes a simple form, for there is only one allowed hopping amplitude between the two Wannier centers of this system.

\begin{equation}\label{eq:molham}
H(k) = 
\begin{pmatrix}
h_z & h_{xy} e^{-i k L} \\
h_{xy} e^{i k L} & -h_z
\end{pmatrix}
\end{equation}

\noindent
with $k$ being the crystal momentum along $\vec{v}$ and $L$ the size of the supercell. Here, $h_{xy}=\sqrt{h_x^2+h_y^2}$. If the band gap $E_g$ of this system is considered fixed, the graph of $\vec{h}(k)$ for its Hamiltonian is a circle (Fig.~\ref{fig:bound1d}b) constrained to lie on a sphere of radius $E_g = \sqrt{h_{xy}^2 + h_z^2}$.   The only degree of freedom available for maximizing the nonlinear response is $h_z$. The extreme cases of $h_z=0$ and $h_z=E_g$ both give no nonlinear response due to reasons of centrosymmetry and vanishing oscillator strength, respectively. We find (methods section~\ref{sec:molbound}) that the optimal value is $h_z = E_g / \sqrt{3}$, which gives 

\begin{equation}\label{eq:molbound}
M <  \frac{\pi e^3}{\hbar} n \left( \frac{\hbar f_{cv}}{mE_g} \right)^{3/2}
\end{equation}

\noindent where $n$ is the number density of molecules and $f_{cv}$ the oscillator strength of the HOMO-LUMO transition.  

We emphasize that the formalism~\cite{Sipe00p5337} used to derive Eqs.~\ref{eq:sc}, \ref{eq:shg} only considers the resonant component of the nonlinear response. In this respect, Eq.~\ref{eq:molbound} is different from the off-resonant SHG bound for molecules proved by Kuzyk~\cite{Kuzyk00p1218}. In using Eqs.~\ref{eq:sc}, \ref{eq:shg}, we are implicitly assuming that broadening of bands caused by phonons, disorder, or many-body effects is less than the band width. In this limit, the amount of broadening (width of $\delta$-functions) in Eqs.~\ref{eq:sc}, \ref{eq:shg} is inconsequential as it does not affect the energy integral in $M = \lvert \int \, \sigma dE \rvert$. In contrast, the expressions for resonant molecular SHG in Ref.~\cite{Kuzyk06p154108} depend on a broadening parameter because they are applicable in the limit where broadening is larger than the band width. Therefore, the bound Eq.~\ref{eq:molbound} proved here should be interpreted as the maximum SHG of almost isolated systems, as the hopping between images tends towards zero. 

Next, we consider the opposite limit of strong hopping between sub-systems. We consider a one dimensional system defined by the Hamiltonian $\vec{h}(k)$ with a fixed band gap $E_g$, and increase the hopping strength along the periodic direction. As the hopping strength increases, the graph of $\vec{h}(k)$ is allowed to change from a circle (Fig.~\ref{fig:bound1d}b) to a path with $\min_k \lvert \vec{h}(k) \rvert = E_g$ (Fig.~\ref{fig:bound1d}c).  As a concrete example, consider the distortion in Fig.~\ref{fig:bound1d}c which changes the graph of $\vec{h}(k)$ from a circle to an ellipse. The dimensions of this ellipse increase with the hopping strength. As this happens, the majority of the nonlinear response is concentrated near the band edges (light colored region in Fig.~\ref{fig:bound1d}c).  In addition, the magnitude of $\vec{h}''(k)$ near the band edge increases, which increases the total amount of nonlinear response, as alluded to above (Eq.~\ref{eq:hhh}). We show (methods section~\ref{sec:1dbound}) that these features are present in general for one-dimensional systems, under the assumptions of finite range hopping (Eq.~\ref{eq:expdecay}) and nondegenerate band minima. For such systems, we derived (methods section~\ref{sec:1dbound}) that the metric for total nonlinear response follows the asymptotic bound

\begin{equation}\label{eq:1dbound}
M < \frac{\pi e^3}{2\hbar} \frac{A}{E_g} \Xi_1(\xi)
\end{equation}

\noindent as $A/E_g\rightarrow\infty$, where $\Xi_1(\xi)= 2 n_1 L^2 \frac{e^{-1/\xi}+e^{-2/\xi}}{(1-e^{-1/\xi})^3}$ is the geometrical factor and $n_1$ is the areal density of these one-dimensional systems. This bound, being proportional to $A/E_g$, is tighter than the general bound ($(A/E_g)^2$, Eq.~\ref{eq:genbound}) in the strong hopping limit (large $A/E_g$). The reason for the different power law in the one-dimensional strong hopping limit can be deduced from Eq.~\ref{eq:hhh}. While $\vec{h}'(k)$ and $\vec{h}''(k)$ are both proportional to $A$ in magnitude, the region of the Brillouin zone which contributes to the nonlinear response is inversely proportional to $A$, leading to the overall linear in $A$ scaling of Eq.~\ref{eq:1dbound}.   
In Fig.~\ref{fig:bound1d}a, we combine the above bounds for the isolated and strong hopping limits to deduce the general behavior for the nonlinear response of a system as a function of hopping strength. At weak hopping between almost isolated systems, the bound is independent of the hopping strength (Eq.~\ref{eq:molbound}), while it is proportional to the hopping strength for large hoppings (Eq.~\ref{eq:1dbound}), implying that delocalized systems have greater potential for large nonlinear responses. This trend is in agreement with theoretical proposals~\cite{Zhou06p2891,Szafruga10p379} and experimental observations in conjugated systems~\cite{Perez-moreno09p5084,Coe10p3496}. With the hopping strength allowed potentially increase without bound, Eq.~\ref{eq:1dbound} suggests that Kuzyk's bound for the SHG of isolated molecules can be broken by sufficient conjugation of molecules.

In summary, we have derived a general upper limit for the shift current and second harmonic generation responses of extended systems, showing that it is controlled by the ratio of the hopping strength to the band gap of the material, as well as being dependent on a geometrical factor. We have separately derived bounds in the strong- and weak-hopping limits, showing that coupling between components tends to increase the amount of nonlinear response. These bounds may be used to guide materials research, by suggesting materials with potentially large responses, or as a screening tool to rule out unfavorable candidates. Besides the design of individual shift current or SHG materials, this work suggests that similar analytical relations may be found for other optical phenomena in solid state materials, such as high order frequency mixing processes, multi-photon absorption, and Raman scattering.

\section{Acknowledgements}
L.Z.T. was supported by the U.S. ONR under Grant N00014-17-1-2574.
A.M.R. was supported by the U.S. Department of Energy, under grant DE-FG02-07ER46431.
Computational support was provided by the HPCMO of the U.S. DOD and the NERSC of the U.S. DOE.

\section{methods}

\subsection{Derivation of general bound}\label{sec:genbound}

The derivatives of the Hamiltonian along direction of current and light polarization $\vec{v}$, in Fourier components, are

\begin{equation}
\vec{h}'(\vec{k}) = \sum_{n_1n_2n_3} \vec{h}_{n_1n_2n_3} 
i (\sum_j R_{j1} v_j n_1 + \sum_j R_{j2} v_j n_2 +\sum_j R_{j3} v_j n_3)
\exp\left(i\vec{k}\cdot(n_1\vec{R}_1+n_2\vec{R}_2+n_3\vec{R}_3)\right)
\end{equation}

\begin{equation}
\vec{h}''(\vec{k}) = \sum_{n_1n_2n_3} \vec{h}_{n_1n_2n_3} 
i^2 (\sum_j R_{j1} v_j n_1 + \sum_j R_{j2} v_j n_2 +\sum_j R_{j3} v_j n_3)^2
\exp\left(i\vec{k}\cdot(n_1\vec{R}_1+n_2\vec{R}_2+n_3\vec{R}_3)\right)
\end{equation}

\noindent where $R_{ji}$ is the $j$-th component of $\vec{R}_i$. For ease of computation, we express the $k$-vectors in units of the reciprocal lattice vectors (crystal coordinates), $\tilde{k}_i=\frac{1}{2\pi}\sum_j R_{ji} k_j$, obtaining

\begin{equation}\label{eq:mcrys}
\begin{aligned}
M=\frac{\pi e^3}{2\hbar} \frac{1}{(2\pi)^3}
\biggr \lvert 
\int d^3\tilde{k} & \frac{(2\pi)^3}{\lvert \det R \rvert}\frac{(i)^3}{\lvert \vec{h} \rvert^3}  
\sum_{n_1 n_2 n_3}\sum_{n_1' n_2' n_3'}\sum_{n_1'' n_2'' n_3''}
(\vec{h}_{n_1''n_2''n_3''} \cdot \vec{h}_{n_1n_2n_3} \times \vec{h}_{n_1'n_2'n_3'} ) \\
& (\tilde{v}_1n_1 +\tilde{v}_2n_2+\tilde{v}_3n_3)
(\tilde{v}_1n_1' +\tilde{v}_2n_2'+\tilde{v}_3n_3')^2 \\
&\exp\left(2\pi i (\tilde{k}_1n_1+\tilde{k}_2n_2+\tilde{k}_3n_3) \right)
\exp\left(2\pi i (\tilde{k}_1n_1'+\tilde{k}_2n_2'+\tilde{k}_3n_3') \right)\\
&\exp\left(2\pi i (\tilde{k}_1n_1''+\tilde{k}_2n_2''+\tilde{k}_3n_3'') \right)
\biggr\rvert
\end{aligned}
\end{equation}

\noindent First, we bound the triple scalar product in Eq.~\ref{eq:mcrys}

\begin{equation}\label{eq:mlevi}
\begin{aligned}
M<\frac{\pi e^3}{2\hbar}
\biggr \lvert 
\int d^3\tilde{k} & \frac{1}{\lvert \det R \rvert}\frac{1}{\lvert \vec{h} \rvert^2} 
\sum_{n_1 n_2 n_3}\sum_{n_1' n_2' n_3'} 
\lvert \vec{h}_{n_1n_2n_3} \rvert \lvert \vec{h}_{n_1'n_2'n_3'} \rvert \\
& (\tilde{v}_1n_1 +\tilde{v}_2n_2+\tilde{v}_3n_3)
(\tilde{v}_1n_1' +\tilde{v}_2n_2'+\tilde{v}_3n_3')^2 \\
&\exp\left(2\pi i (\tilde{k}_1n_1+\tilde{k}_2n_2+\tilde{k}_3n_3) \right)
\exp\left(2\pi i (\tilde{k}_1n_1'+\tilde{k}_2n_2'+\tilde{k}_3n_3') \right)
\biggr \rvert 
\end{aligned}
\end{equation}

\noindent Next, we let $f = \frac{1}{\lvert \vec{h} \rvert^2}$, and bound the Fourier components of this quantity. By Parseval's theorem, we have

\begin{equation}\label{eq:parseval}
\lvert f_{n_1n_2n_3} \rvert^2 < \sum_{n_1n_2n_3} \lvert f_{n_1n_2n_3} \rvert^2  = \int d^3 \tilde{k} \, \lvert f(\tilde{k}) \rvert^2 < \frac{1}{E_g^4}
\end{equation}

\noindent As a consequence, Eq.~\ref{eq:mlevi} becomes

\begin{equation}\label{eq:mrecip}
\begin{aligned}
M<\frac{\pi e^3}{2\hbar}
& \frac{1}{\lvert \det R \rvert} \frac{1}{E_g^2}
\sum_{n_1 n_2 n_3}\sum_{n_1' n_2' n_3'} 
\lvert \vec{h}_{n_1n_2n_3} \rvert \lvert \vec{h}_{n_1'n_2'n_3'} \rvert \\
& (\tilde{v}_1n_1 +\tilde{v}_2n_2+\tilde{v}_3n_3)
(\tilde{v}_1n_1' +\tilde{v}_2n_2'+\tilde{v}_3n_3')^2 
\end{aligned}
\end{equation}

\noindent Inserting the bounds Eq.~\ref{eq:expdecay} and performing the sums over $n_i, n_i'$ yields Eqs.~\ref{eq:genbound}, \ref{eq:geomfac} of the main text.

\subsection{Derivation of bound in the weak hopping limit}\label{sec:molbound}

For the Hamiltonian in Eq.~\ref{eq:molham}, Eq.~\ref{eq:hhh} gives 

\begin{equation}
M = \frac{\pi e^3}{2\hbar} \frac{L^3}{V} \frac{h_{xy}^2h_z}{( h_{xy}^2+h_z^2 )^{3/2}}
\end{equation}

\noindent where $V$ is the volume of the supercell. For fixed band gap $E_g = \lvert h\rvert$, this is maximized at $h_z = \lvert h\rvert/\sqrt{3}$. To rewrite this in terms of the oscillator strengths, we use $f_{cv} = \frac{2m\omega_{cv}}{\hbar} \lvert r_{cv} \rvert^2$, and

\begin{equation}\label{eq:r12}
\lvert r_{cv} \rvert^2 = \frac{
(h_z \frac{d\lvert h\rvert}{dk}-\lvert h\rvert \frac{dh_z}{dk})^2
+(h_x \frac{dh_y}{d_k} - h_y \frac{dh_x}{d_k})^2}
{4 \lvert h\rvert^2 (\lvert h\rvert^2-h_z^2)}
\end{equation}

\noindent as derived in~\cite{Fregoso16preprint}. At the optimum point, we have $\lvert r_{cv} \rvert^2=L^2/6$, which results in the bound Eq.~\ref{eq:molbound}. 

\subsection{Derivation of bound in the strong hopping limit}\label{sec:1dbound}

To derive Eq.~\ref{eq:1dbound}, we start with some arbitrary fixed Hamiltionian $\vec{h}_{\text{fix}}(k)$ and add an adjustable correction, so that $\vec{h}(k) = \vec{h}_{\text{fix}}(k) +\lambda \vec{\Delta}(k)$. We are interested here in the strong hopping limit of large $\lambda$.  We assume that the Fourier components are exponentially bounded, as in Eq.~\ref{eq:expdecay}: 

\begin{equation}
\begin{aligned}
\lvert \vec{h}_{\text{fix},n} \rvert &< A_{\text{fix}} e^{-n/\xi} \\
\lvert \vec{h}_{n} \rvert &< A e^{-n/\xi}
\end{aligned}
\end{equation}

\noindent where $\vec{h}_{\text{fix}}(k) = \sum_n \vec{h}_{\text{fix},n} e^{inkL}$, $\vec{h}(k) = \sum_n \vec{h}_{n} e^{inkL}$, and $\vec{\Delta}(k) = \sum_n \vec{\Delta}_{n} e^{inkL}$. Since we are interested in the limit $A\rightarrow\infty$, we generally have $A>A_{\text{fix}}$. The triangle inequality then implies that 

\begin{equation}
\lambda \lvert \vec{\Delta}_n \rvert < 2A e^{-n/\xi}
\end{equation}

Furthermore, we assume that the correction does not change the band gap of the system: $\vec{\Delta}(k=0)=0$, and that the band gap occurs at only a single point in the Brillouin zone. Here, the band gap location is taken to be at $k=0$ without loss of generality.  Apart from these conditions, the form of the correction is otherwise not constrained. 

We write the metric for the integrated nonlinear response as (Eq.~\ref{eq:hhh})

\begin{equation}\label{eq:hhhcorr}
M = \frac{\pi e^3}{2\hbar} n_1 \left \lvert
\int \frac{dk}{2\pi} 
\frac{\hat{h}(k) \cdot 
(\vec{h}_{\text{fix}}'(k) + \lambda \vec{\Delta}'(k) )\times (\vec{h}_{\text{fix}}''(k)+ \lambda \vec{\Delta}''(k) )}{\lvert \vec{h}_{\text{fix}}(k) + \lambda \vec{\Delta}(k) \rvert^2 } 
\right \rvert
\end{equation}

\noindent Performing a Taylor expansion about $k=0$,

\begin{equation}
\begin{aligned}
\vec{h}_{\text{fix}}(k) =& \, \vec{h}_{\text{fix}}(0) + \vec{h}_{\text{fix}}'(0) k + \frac{1}{2} \vec{h}_{\text{fix}}''(0) k^2 \\
\vec{\Delta}(k) =& \, \vec{\Delta}'(0) k + \frac{1}{2} \vec{\Delta}''(0) k^2 
\end{aligned}
\end{equation}

\noindent we see that the factor $1/\lvert h\rvert^2$ approaches a $\delta$-function as $\lambda\rightarrow\infty$

\begin{equation}\label{eq:deltafn}
\lim_{\lambda\rightarrow\infty} \frac{1}{\lvert \vec{h}_{\text{fix}}(k) + \lambda \vec{\Delta}(k) \rvert^2 }
= \lim_{\lambda\rightarrow\infty} \frac{1}{\lvert \vec{h}_{\text{fix}}(0) \rvert^2 + (\lambda \vec{\Delta}'(0))^2 k^2}
= \frac{\pi}{\lambda E_g \lvert \vec{\Delta}'(0)\rvert} \delta(k)
\end{equation}

\noindent The $\lambda^2$ terms in the numerator of Eq.~\ref{eq:hhhcorr} dominate as $\lambda\rightarrow\infty$, which yields

\begin{equation}\label{eq:1dboundpre}
M < \frac{\pi e^3}{2\hbar} n_1 \frac{\lambda\lvert \vec{\Delta}''(0)\rvert }{2 E_g}
\end{equation}

\noindent The second derivative $\lambda\lvert \vec{\Delta}''(0)\rvert$ is bounded by

\begin{equation}\label{eq:2derivbound}
\lambda\lvert \vec{\Delta}''(0)\rvert < \sum_n n^2L^2 \lambda\lvert \vec{\Delta}(0)\rvert < 2A \sum_n n^2L^2 e^{-n/\xi}
\end{equation}

\noindent Combining Eqs.~\ref{eq:1dboundpre} and \ref{eq:2derivbound} results in Eq.~\ref{eq:1dbound} of the main text. Finally, we note that the degenerate case $\vec{\Delta}'(0)=0$ in Eq.~\ref{eq:deltafn} does not affect this bound. In this case, Eq.~\ref{eq:deltafn} becomes

\begin{equation}
\lim_{\lambda\rightarrow\infty} \frac{1}{\lvert \vec{h}_{\text{fix}}(k) + \lambda \vec{\Delta}(k) \rvert^2 }
= \lim_{\lambda\rightarrow\infty} \frac{1}{\lvert \vec{h}_{\text{fix}}(0) \rvert^2 + (\lambda \vec{\Delta}''(0)\cdot \vec{h}_{\text{fix}}(0)) k^2}
=  \frac{\pi}{E_g \sqrt{\lambda \lvert \vec{\Delta}''(0)\cdot \vec{h}_{\text{fix}}(0)\rvert}} \delta(k)
\end{equation}

\noindent while the numerator of Eq.~\ref{eq:hhhcorr} scales as $\hat{h}(k) \cdot \vec{h}'_{\text{fix}}(k) \times \lambda \vec{\Delta}''(k)$, leading to an overall scaling of $\sqrt{\lambda}$, which is of subleading order compared to Eq.~\ref{eq:1dboundpre}.

\clearpage

\clearpage

\bibliography{rappecites,newcites}

\clearpage

\end{document}